# Planar Nernst and Hall effect in patterned ultrathin film of $La_{0.7}Sr_{0.3}MnO_3$

Himanshu Sharma[1*], Hanuman Bana[2], Ashwin Tulapurkar[2], C. V. Tomy[1]

[1]Department of Physics, Indian Institute of Technology Bombay, Powai, Mumbai – 400 076, India.
[2]Department of Electrical Engineering, Indian Institute of Technology Bombay, Powai, Mumbai – 400 076, India.
E-mail: * himsharma@iitb.ac.in

We present the observation of *transverse thermopower*, or *planar Nernst Effect* (PNE), and *planar Hall Effect* (PHE) in patterned ultrathin film (~ 8 nm) of $La_{0.7}Sr_{0.3}MnO_3$ (LSMO). For PNE, a heat current was applied in-plane of the film (by creating a temperature gradient $\nabla T$) whereas an in-plane current of 100 µA was applied for measuring PHE. Even for a temperature difference of $\Delta T = 5$ K the planar Nernst effect in LCMO ultrathin film (~ 8 nm) shows a coercivity of ~ 3 Oe and confirms the ferromagnetic nature of the thin film. The angular dependence of transverse voltage in PNE shows a four-fold $\sin\theta \cos\theta$ dependence symmetry at 250 K, which is consistent with the planar Hall effect measurement.

## 1. Introduction

In addition to the planar Hall effect (PHE), the thermoelectric and spin caloritronic effects, where a sample develops a transverse voltage in response to a thermal gradient ($\nabla T$) applied in-plane of the sample, have attracted detailed attention due to its potential applications based on the generation and detection of spin current [1,2]. The contribution of planar Nernst effect (PNE), or transverse thermopower in spin Seebeck effect (SSE) has been observed in many materials [3,4,5]. Nernst effect gives these materials an advantage for application and increases the usability of these materials in spin-based devices [12, 64]. Tremendous research efforts have been devoted in spin caloritronics to investigate the interplay of spin and thermoelectric transport phenomena [1,6,7,8,9]. $La_{0.7}Sr_{0.3}MnO_3$ (LSMO) is an important material for spintronics due to its large spin-polarization [3]. We have shown previously that the ferromagnetic transition temperature ($T_C$) and saturation magnetization ($M_s$) of LSMO films are sensitive to the film thickness [10,11]. It was found that the $T_C$ as well as the $M_s$ decreases with decreasing the film thickness [11]. In addition, the thickness of the film has to be optimized for the manipulation of magnetization by spin current, as the spin transfer torque effect is inversely proportional to the thickness [12,13]. Since the minimum thickness needed for retaining the $T_C$ above room temperature and the $M_s$ almost the same as the bulk value is 8 nm in thin LSMO films. We have prepared the LSMO film of 8 nm thickness for the studies described in this paper.

## 2. Experimental techniques

We have deposited thin films of LSMO with film thickness down to 8 nm on STO (001) substrates (see Fig. 1(a)) by the pulsed laser deposition (PLD) method using a KrF-pulsed LASER source having a wavelength of 248 nm and energy density [10] of ~ 2 Jcm$^{-2}$. During the deposition, the oxygen pressure and the substrate temperature were kept at $3.5\times10^{-1}$ mbar and 750ºC, respectively.

For measuring the planar Hall effect (PHE), these LSMO thin films were converted into patterned Hall bar structures, as shown in Fig. 1(b), by using the optical lithography and ion-milling process. The current and voltage channels are marked as $(+ I, − I)$ and $V_{PHE}$ (see Fig. 1(b)). Fig. 1(c) shows the SEM images of one of the LSMO channels. The typical lateral sizes of the LSMO channels are 100 µm × 2.4 mm, whereas the distance between the two voltage probes is 100 µm. In order to convert this structure for planar Nernst effect (PNE) measurements, the Hall bar structure was mounted on a platform as shown in Fig. 1(d).

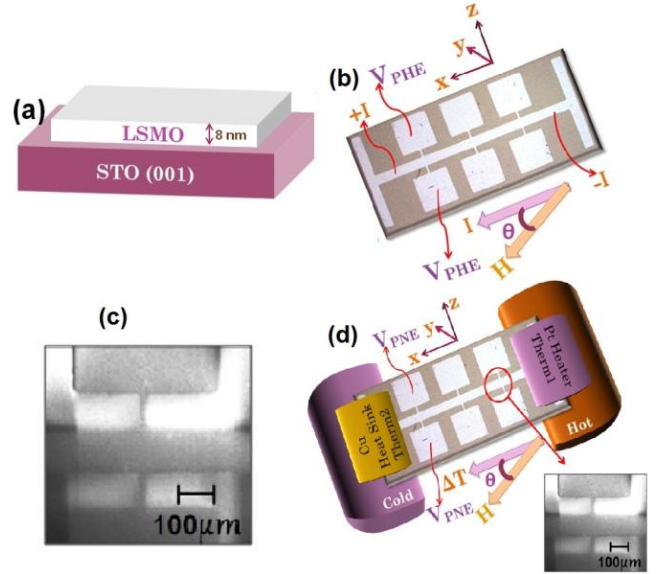

**Fig. 1.** Schematic of (a) LSMO thin film, (b) Hall bar structure using optical microscope image, (c) SEM image of one of the LSMO channels and (d) setup for Planar Nernst effect measurement.

A Pt (Pt-1000) heater was used at one end (hot) and a Cu heat sink (cold) was used at the other end of the film to obtain the required temperature gradient. Temperatures at the hot and cold ends were measured using individual Pt thermometers. A cryostat of physical property measurement system (PPMS-Quantum Design Inc.) was used for obtaining temperature and magnetic field variations. For making the angle dependent measurements by rotating the sample continuously in the film plane during the measurements, the Hall bar structure was mounted on a sample platform of an in-plane rotator (Horizontal rotator-Quantum Design Inc., USA). This allowed us to change the relative angle from 0° to 360° and then backwards between *I* and *H* for PHE measurements or between ∇T and *H* for PNE measurements. For the planar Hall effect measurements, an in-plane dc current of 100 μA was applied along the x-axis to record the corresponding transverse output voltage along the y-direction. For the planar Nernst effect measurements, an in-plane temperature gradient ($\nabla T_x$) is applied along the x-axis using the Pt-heater and a Source Meter (Keithley - 2602A). The transverse output voltage was measured using a nanovoltmeter (Keithley - 2182A). The transverse voltages (with or without rotation) in both the cases were recorded at different temperatures and in varying magnetic fields.

## 3. Results and discussions

Figure 2(a) shows the XRD pattern for the $La_{0.7}Sr_{0.3}MnO_3$ (LSMO) thin film on $SrTiO_3$ (STO) substrate. From the XRD data in the expanded form (inset of Fig. 2(a)), we can identify a strong (002) peak of LSMO quite close to the STO (002) peak, which confirms the epitaxial growth of the film. The lattice mismatch between the STO (a = 3.905 Å) and the LSMO (a = 3.871 Å) is only ~ 1 %. Because of such a small lattice mismatch, we could obtain a uniform grain formation of the LSMO on the STO surface.

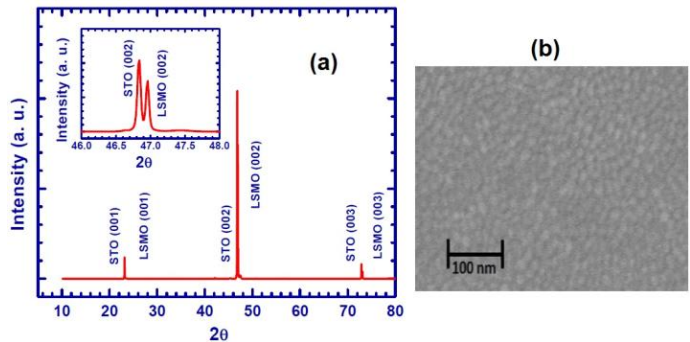

**Fig. 2.** (a) X-ray diffraction patterns of the LSMO thin film. Inset shows the pattern in the expanded scale to highlight the LSMO (002) peak. (b) SEM image of the surface of LSMO thin film.

The SEM image, shown in Fig. 2(b), confirms the continuity in the film (no pin holes and cracks).

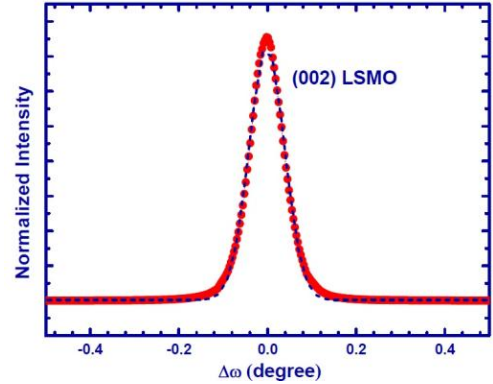

**Fig. 3.** High resolution rocking curve (ω-scan) profile around the (002) crystallographic peak for the LSMO film grown on STO substrate.

The crystalline quality of the LSMO film was checked by measuring the full-width-half-maximum (FWHM) of the rocking curve (ω-scan) as shown in Fig. 3. The FWHM value of the ω-scan around the (002) diffraction peak of the LSMO was found to be ~ 0.046 deg, which confirms the high quality crystalline structure of the film.

*3.1 Magnetization measurements*

We first characterized the magnetic properties of the prepared LSMO thin film by measuring the in-plane magnetization as a function of temperature and applied magnetic field using a SQUID-VSM (Quantum Design, USA). The temperature dependence of the field-cooled (FC) magnetization of the thin film with an applied magnetic field of 1 kOe is shown in Fig. 4.

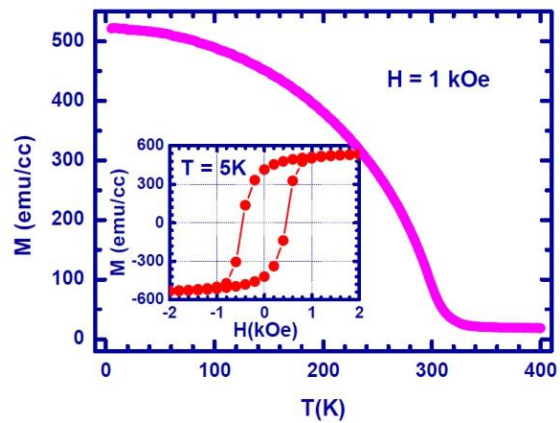

**Fig. 4.** In-plane FC magnetization as a function of temperature (*T*) in an applied field of 1 kOe for 8 nm LSMO/STO thin film. Inset shows the variation of in-plane magnetization as a function of field *H* at 5 K.

For the thin film of 8 nm, the ferromagnetic transition temperature ($T_C$) is observed above room temperature (~ 325 K). Inset of Fig. 4 shows the magnetization as a function of magnetic field at 5 K; the saturation behaviour in magnetization confirms the ferromagnetic nature of this film.

*3.2 Planar Hall Effect Measurements*

Initially, the transverse voltage ($V_{xy}$) in zero applied magnetic field was measured as a function of temperature by passing an in-plane current of 100 μA, which is shown as one of the curves in Fig. 5(a). Ideally, if the transverse voltage leads were perfectly symmetrical, one would have observed zero voltage since the applied magnetic field is zero. Thus the observed transverse voltage is not a Hall voltage. The observed behaviour of $V_{xy}$ almost resemble the normal resistance variation with a peak at the Insulator to metal transition, as shown in Fig. 5(b) where the longitudinal voltage ($V_{xx}$) is plotted as a function of temperature. Temperature variation of $V_{xy}$ and $V_{xx}$ as a function of temperature at various applied magnetic fields is plotted in Fig. 5(a) and (b). The field dependence behaviour is certainly different indicating a clear contribution of planar Hall voltage in $V_{xy}$. This is even more evident in Fig. 5(c) and (d) where we have plotted the differences $\Delta V_{xy}$ (= $V_{xy}(0) - V_{xy}(H)$) and $\Delta V_{xx}$ (= $V_{xx}(0) - V_{xx}(H)$) by subtracting zero field contribution.

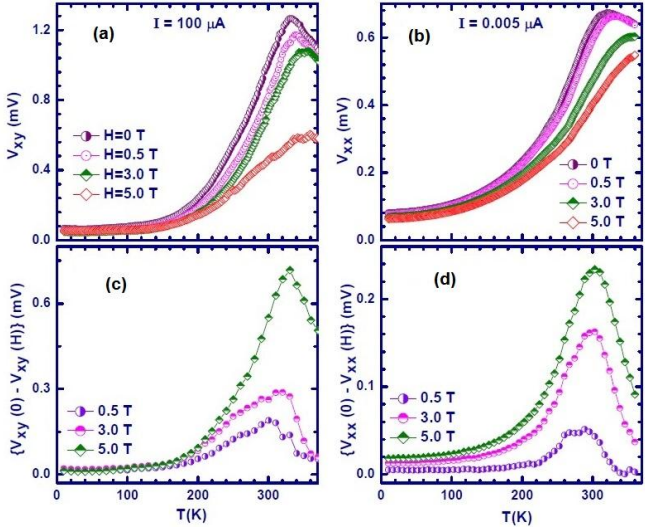

**Fig. 5.** Temperature dependence of (a) $V_{xy}$ and (b) $V_{xx}$ for LSMO/STO ultrathin film of 8 nm in applied magnetic field of 0.5 T, 1 T, 3T and 5T at θ = 45°, (c) $\Delta V_{xy}$ and (d) $\Delta V_{xx}$ as a function of temperature in different applied magnetic fields.

This observation of peak in PHE near the ferromagnetic transition is consistent with the previously reported results in manganite [14,15]. Such peculiar peaking near the insulator-metal transition in manganites has been explained [14] based on the microscopic inhomogeneities across the film.

Angular dependence of the longitudinal voltage ($V_{xx}$) and the transverse voltage ($V_{xy}$) was then measured at 250 K in an applied magnetic field of 0.5 T by varying the angle θ from 0° to 360°, where θ = 0° corresponds to the configuration of *H* perpendicular to the current (*I*).

In Fig. 6(a), we have shown the angular dependence of $V_{xx}$ whereas the same dependence of $V_{xy}$ is given in Fig. 6(b). Here, the constant part in $V_{xy}$, which is contributed from $V_{xx}$ was not removed, as it will not affect the angular dependence of the transverse voltage.

The longitudinal voltage ($V_{xx}$) and the transverse voltage ($V_{xy}$) (in the high-field limit) can be phenomenologically described by the equations [16]:

$$V_{xx} = A_0 + A_1 \cos^2\theta \quad ------ \quad (1)$$
$$V_{xy} = A_2 + A_3 \sin\theta \cos\theta \quad ------ \quad (2)$$

Where, θ is the angle between magnetic field and the applied current and the coefficients $A_i$ depend on the magnetic field and the temperature at which measurements are performed.

The longitudinal voltage ($V_{xx}$) shows an angular dependence with a $\cos^2\theta$ variation consistent with Eq. (1). The solid (red) line in Fig. 6(a) shows the fit, which matches well with the data.

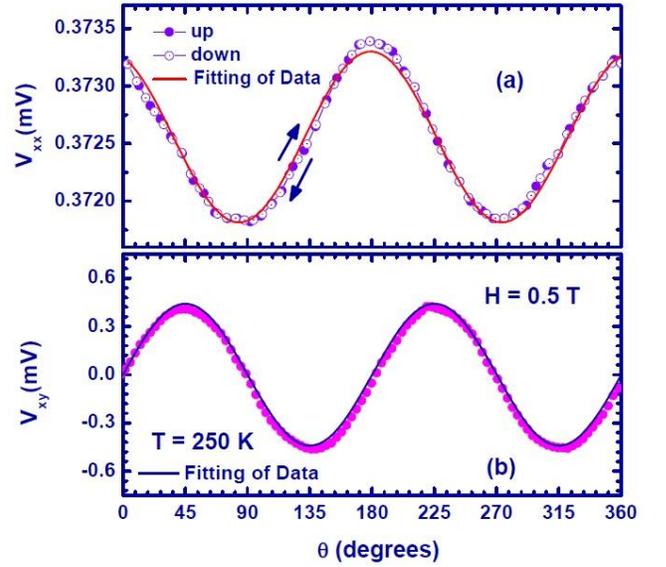

**Fig. 6.** Angular dependence of (a) longitudinal Voltage ($V_{xx}$) and (b) transverse voltage ($V_{xy}$) for LSMO/STO ultrathin film of 8 nm in applied magnetic field of 0.5 T at 250 K. The line shows the fit to $\cos^2\theta$ and $\sin\theta \cos\theta$, respectively.

The angular dependence of the transverse voltage ($V_{xy}$) was fitted to Eq. (2) and is shown as the blue line in Fig. 6(b). This change in symmetry from the $\cos^2\theta$ dependence in $V_{xx}$ to the $\sin\theta \cos\theta$ dependence symmetry in $V_{xy}$ confirms the dominating nature of the planar Hall effect in transverse voltage ($V_{xy}$). In addition, there should be shift of 45° between $V_{xx}$ (typically the AMR) and $V_{xy}$ (PHE), which is what we have observed in our results, reconfirming the contribution of PHE in $V_{xy}$.

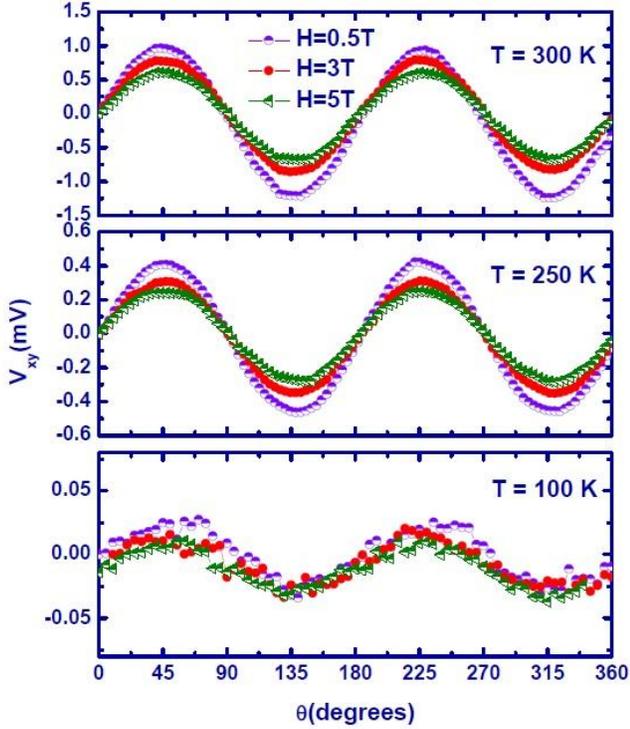

**Fig. 7.** Angular dependence of the transverse voltage ($V_{xy}$) for LSMO/STO ultrathin film of 8 nm in applied magnetic field of 0.5 T, 3 T and 5 T at 100 K, 250 K and 300 K.

We have further measured the effect of magnetic field on the angular dependence of $V_{xy}$ at 100 K, 250 K and 300 K. Figure 7 shows the angular dependence of $V_{xy}$ in applied magnetic fields of 0.5 T, 3 T and 5 T. It is evident that, for all temperatures and magnetic fields, the angular dependence of PHE shows the $\sin\theta\cos\theta$ dependence symmetry with varied amplitude and the curves can all be well described by Eq. 2 (See Fig. 7). The observation of a significant four-fold oscillation in PHE as well as the decrease in magnitude with increasing magnetic field is consistent with results reported for manganite thin films on STO by other groups [15,16].

*3.3 Planar Nernst Effect Measurements*

Planar Nernst effect (PNE) is a complimentary measurement tool to planar Hall effect, and both originate from the spin-orbit interactions, there will be a correspondence between the two measurements in the same material. In the next experiment to study the Planar Nernst effect (PNE), we created a temperature difference across the LSMO thin film by using a Pt-heater mounted on hot end of the sample and measured the transverse voltage between the two pads using a nanovoltmeter. The angular dependence of transverse voltages is recorded at 250 K and in a magnetic field of 0.5 T for a temperature difference of $\Delta T = 5$ K as shown in Fig. 8(a).

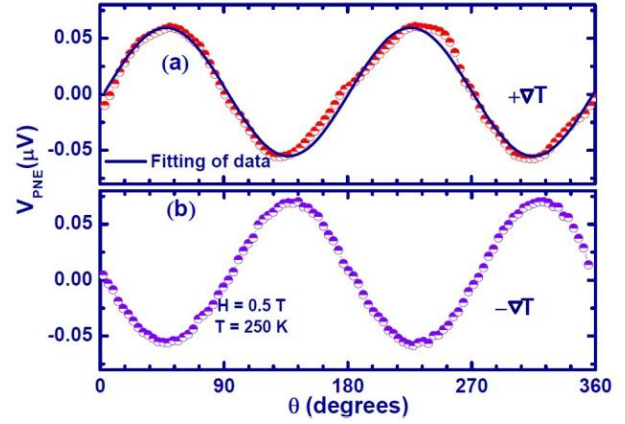

**Fig. 8.** Angular dependence of the Planer Nernst voltage in applied magnetic field of 0.5 T at 250 K with applied in-plane temperature gradient of (a) $+\nabla T_x$ and (b) $-\nabla T_x$.

The true nature of the PNE was confirmed by reversing temperature gradient direction and measuring the angular dependence of PNE. Figure 8(b) shows the data when the hot and the cold ends were reversed, thus introducing a sign change in $\nabla T_x$. The angular dependence of PNE resembles its magnetic counterpart, PHE, thus confirming their common physical origin. With the in-plane thermal gradient, the transverse PNE voltage depends on the the magnetization M of the ferromagnet and the magnitude of the thermal gradient, given by the equation [3]:

$$V_{PNE} \propto |M|^2 \, |\nabla T_x| \, \sin\theta\cos\theta \quad \text{-------} \quad (3)$$

Where, $\theta$ is the angle between $\nabla T_x$ and M. The angular dependence of transverse voltage $V_{PNE}$ shows a $\sin\theta\cos\theta$ dependence symmetry at 250 K, which is consistent with the planar Hall effect measurement and is in good agreement with the predictions of Eq. 3, for Planar Nernst effect (in Fig. 8(a), the line is a fit to Eq. 3).

In order to investigate the effect of magnetic field on PNE, we have measured the planar Nernst voltage or the transverse voltage ($V_{PNE}$) in response to a thermal gradient ($\nabla T_x$) as a function of applied field at 250 K. We have chosen two orientations, $\theta = 45°$ where the $V_{PNE}$ is maximum (corresponding to the easy axis of magnetization) and $\theta = 135°$, where $V_{PNE}$ is minimum, corresponding to the hard axis of magnetization). The absolute value of Planer Nernst voltage ($V_{PNE}$) increases initially with the applied magnetic field, until saturation is reached at $H_{sat} = \sim 15$ Oe.

The initial linear and then the saturation behaviour is the typical variation of the $V_{PNE}$ with magnetic field observed in the ferromagnetic manganites. Due to a constant thermoelectric offset voltage, the curves are not centered at H = 0. The coercivity (~ 3 Oe) observed (see Fig. 9) for both the orientations ($\theta = 45°$ and $\theta = 135°$) confirm the ferromagnetic nature of this film at 250 K.

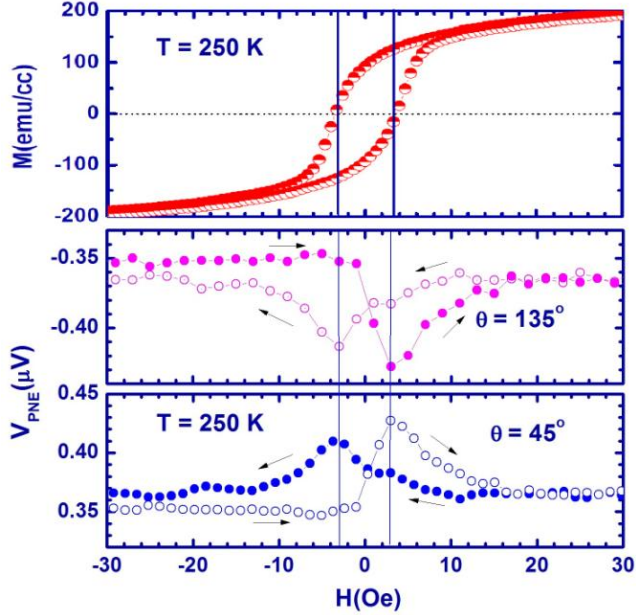

**Fig. 9.** Magnetization as a function of field *H* and $V_{PNE}$ as a function of applied magnetic field at 250 K at $\theta = 45°$ and $\theta = 135°$.

The results obtained in this paper bring in interesting aspects regarding the sensitivity to thermal gradient, resulting in observable voltage changes under small temperature differences in planar Nernst effect (PNE). On the other hand, planar Hall effect (PHE) shows extreme sensitivity to magnetic fields. LSMO ultrathin (8 nm) film, with tensile strain, shows fourfold $\sin\theta \cos\theta$ angular dependence of transverse or planar Hall voltage ($V_{xy}$) and planar Nernst voltages ($V_{PNE}$). This implies that only small magnetic fields are needed to obtain saturation ($H_{sat}$ = 15 Oe) in $V_{PNE}$ with the tensile strain in thinner or ultrathin films. The sensitivity of PNE to $\nabla T$ could be utilized to sense small temperature differences across nanostructures. Additionally, a low magnetic field sensing devices can be fabricated by using the sensitivity of PHE to magnetic field. It will be interesting to see whether such a change in magnetization and $V_{PNE}$ can also be observed with high compressive strain in order to fabricate high field or low field thermopower [9] devices by designing these materials with high or low tensile strain and high or low compressive strain.

## 4. Conclusion

In summary, we have investigated planar Hall effect (PHE) in patterned ultrathin film (8 nm) of $La_{0.7}Sr_{0.3}MnO_3$ (LSMO) with in-plane current of 100 µA. Further, planar Nernst effect (PNE) in the same patterned ultrathin film (8 nm) of LSMO driven by heat current applied in-plane of the film is also investigated. It is observed that results observed by both the PHE and PNE are consistent. We found that the coercivity (3 Oe) observed (See Fig. 9) for both orientations ($\theta = 45°$ and $\theta = 135°$), which confirm the ferromagnetic nature of this film at 250 K. Additionally, a low magnetic field sensing device can be fabricated by using the sensitivity of PHE to magnetic field and the sensitivity of PNE to $\nabla T_x$ could be utilized to sense temperature differences across the microstructures.

## Acknowledgments

We are grateful for the availability of the Institute central facility (SQUID-VSM) in the Department of Physics and IITB Nanofabrication facility (IITBNF) in the Department of Electrical Engineering and Centre of Excellence in Nanoelectronics (CEN), Indian Institute of Technology Bombay.